
\newif\ifapj
\ifapj
	\magnification=\magstep1
\else
	\magnification=\magstephalf
\fi
\input epsf
\epsfverbosetrue
\font\caps=cmcsc10
\font\bb=cmbx12
\tolerance=10000 \pretolerance=5000
\looseness=-3
\widowpenalty 2000
\clubpenalty 2000
\displaywidowpenalty 2000
\overfullrule 0pt

\ifapj\baselineskip 24pt
 \else \baselineskip=15truept  
\fi
\parindent=2.5em        
\def\ub{\underbar}
\def\hi{\par\noindent \hangindent=2.5em}
\def\ls{\vskip 12.045pt}   
\def\ni{\noindent}        
\def\et{{\it et\thinspace al.}\ }    
\def\eg{{\it e.~g.}\ }
\def\kms{km\thinspace s$^{-1}$ }     
\def\deg{\ifmmode^\circ\else$^\circ$\fi}    

\def\arcs{\ifmmode {'' }\else $'' $\fi}     
\def\arcm{\ifmmode {' }\else $' $\fi}     
\def\buildrel#1\over#2{\mathrel{\mathop{\null#2}\limits^{#1}}}
\def\mper{\ifmmode \buildrel m\over . \else $\buildrel m\over .$\fi}
\def\hper{\ifmmode \rlap.^{h}\else $\rlap{.}^h$\fi}
\def\sper{\ifmmode \rlap.^{s}\else $\rlap{.}^s$\fi}
\def\arcsper{\ifmmode \rlap.{' }\else $\rlap{.}' $\fi}
\def\arcmper{\ifmmode \rlap.{'' }\else $\rlap{.}'' $\fi}


\def\aj{{AJ}, }  
\def\apj{{ApJ}, }  
\def\apjs{{ApJSup}, }  
\def\apjl{{ApJLett}, } 
\def\mn{{MNRAS}, }      
%
%

\def\spose#1{\hbox to 0pt{#1\hss}}
\def\lta{\mathrel{\spose{\lower 3pt\hbox{$\mathchar"218$}}
     \raise 2.0pt\hbox{$\mathchar"13C$}}}
\def\gta{\mathrel{\spose{\lower 3pt\hbox{$\mathchar"218$}}
     \raise 2.0pt\hbox{$\mathchar"13E$}}}

\def\hmpc{$h^{-1}$\thinspace Mpc}

\ifapj\vglue 29.10pt \fi %

\ifapj
\centerline{\bb MAPPING MODERATE REDSHIFT CLUSTERS}
\else
\centerline{{\bb MAPPING MODERATE REDSHIFT CLUSTERS}\footnote{$^\dagger$}
	{\tt Submitted to the J. R. A. S. C. July 5, 1993}}\fi
\medskip
\centerline{\bf R. G. Carlberg$^{1,12}$, H. K. C. Yee$^1$, Erica Ellingson$^2$,
	C. J. Pritchet$^3$,}
\centerline{\bf Roberto Abraham$^4$, Tammy Smecker-Hane$^4$, J. R. Bond$^5$,}
\centerline{\bf H. M. P. Couchman$^6$, D. Crabtree$^4$,
	D. Crampton$^4$, Tim Davidge$^{11}$, D. Durand$^4$,}
\centerline{\bf S. Eales$^1$, F. D. A. Hartwick$^3$, J. E. Hesser$^4$,
	J. B. Hutchings$^4$,}
\centerline{\bf N. Kaiser$^5$, C. Mendes de Oliveira$^7$, S. T. Myers$^8$,
	J. B. Oke$^4$,}
\centerline{\bf M. A. Rigler$^4$, D. Schade$^9$, and M. West$^{10}$}
\bigskip
\noindent
$^1$Department of Astronomy, University of Toronto,
$^2$Department of Astronomy, University of Colorado,
$^3$Department of Physics and Astronomy, University of Victoria,
$^4$Dominion Astrophysical Observatory, National Research Council,
$^5$Canadian Insitute for Theoretical Astrophysics, University of Toronto,
$^6$Department of Astronomy, University of Western Ontario,
$^7$Institut d'Astrophysique, Paris,
$^8$Department of Astronomy, California Institute of Technology,
$^9$Institute of Astronomy, Cambridge University,
$^{10}$Sterrewacht, Leiden University,
$^{11}$Gemini Project Office, University of British Columbia,
$^{12}$Department of Astronomy, University of Washington

\bigskip
\ifapj
\bigskip
\centerline{received: $\underline{\hbox to 6truecm{July 7, 1993\hfill}}$}
\vskip 0.5truecm
\centerline{accepted: $\underline{\hbox to 6truecm{\hphantom\null}}$}
\vfill\eject
\fi

\centerline{ABSTRACT}

To test whether clusters have rising mass to light ratios at large
radii and to estimate the amplitude of the density fluctuation
spectrum on the scale of 10\hmpc\ the Canadian Network for
Observational Cosmology (CNOC) cluster collaboration is obtaining
Multi-Object Spectrograph velocities and two colour photometry for a
sample of $\simeq$1000 cluster galaxies and $\simeq$2000 field
galaxies in a 5\hmpc\ neighborhood of high X-ray luminosity clusters
at $z\simeq 0.3$.  X-ray selection of the cluster sample picks out
objects on the basis of the depth of their potential well depth, and
is insensitive to both projection effects and galaxy biases.  The
galaxy dataset, with selection controlled using automated photometry,
automated mask design, spectral selection modeling, and accurate
velocities, will be the best available at any redshift for our tests.
Measuring mass-to-light ratios (M/L) at large radii requires a
statistical removal of field galaxies projected into the redshift
space of the cluster.  This can be done relatively accurately for
moderate redshift clusters, using the large number of foreground and
background galaxies.  The CNOC sample also provides a sensitive test
of the density fluctuation spectrum on cluster scales.  The 30 or so
redshifts available in each of three clusters gives an average
velocity dispersion of about 1000 \kms which indicates that the
$\sigma_8$ normalization parameter is in the range of $0.6\lta
\sigma_8\lta 0.9$.


\vfill\eject

\ni\ub{1. INTRODUCTION}

\ls
The mean mass density of the universe is a fundamental
cosmological parameter for which there are two relatively
distinct values implied by observations at two
different length scales. Large scale streaming velocities find
that $\Omega^{0.6}/b\simeq1$ (Bertschinger \et 1990, Kaiser \et 1991,
Strauss \et 1992),
where $b$ is the poorly known bias parameter. For
$b\simeq1$ the streaming velocities are consistent with $\Omega=1$,
as predicted
by inflationary cosmology (Guth 1981, Bardeen \et 1983).
Measurements from somewhat smaller scales,
where virialized motions dominate, favor $\Omega\simeq0.2$
(\eg Kent and Gunn 1982,
Davis and Peebles 1983, Bean \et 1983). Taken together
the two measurements suggest that
large clusters of galaxies may have dark matter halos
that extend far beyond the bulk of their visible
galaxy distribution. Furthermore, there is evidence from
n-body simulations that a
segregation of galaxies and dark matter can develop
even within the dissipationless clustering that leads
to the buildup of a large cluster (West and Richstone 1988,
Carlberg and Couchman 1989,
Carlberg \et 1990,
Carlberg and Dubinski 1991). Individually the galaxies accurately measure
the mass inside their orbits, but their distribution as a population
can be more centrally concentrated than the total mass distribution.
The primary goal of our observations is to establish whether
the mass to light ratio of clusters rises beyond
the virial radius.

At the present time the best data for measuring
cluster masses at large radii
come from galaxy redshifts. The difficulty in interpreting
a redshift map is that it is impossible to distinguish between
a cluster galaxy orbiting with a line-of-sight velocity of, say, 2000 \kms
and a field galaxy with no peculiar velocity situated 20\hmpc\ from
the cluster. The normal procedure for virial mass
measurements is to terminate the analysis
where the projected cluster blends into the field.
This paper presents a straightforward method which treats
the field galaxies as an asset which is used to  remove statistically
the background contamination in the cluster redshift map.
For samples of a few thousand galaxies, spread over about 10
clusters, the method is shown to test reliably for the presence
or absence of extended mass around clusters, as expected in
$\Omega=1$ or $\Omega=0.2$ universes, respectively.

The statistical requirements for measuring the
mass profile motivate the
design of our observations.
A critical assumption of the mass estimation method
is that the cluster has no nonspherical structure beyond linear gradients
in density front to back and side to side. However, clusters
are known to be triaxial, and are embedded in a rich
mixture of surrounding large scale structure
and have the further difficulty that the
sampling efficiency is likely to vary over large angles on
the sky, such as the
$30^\circ$ spanned by the Coma data. The cluster asymmetries can
be averaged out using a small sample.
Clusters at moderate redshifts, $z\simeq0.3$,
can be selected from X-ray surveys, the galaxies within them
can be sampled with good statistical control,
and there are sufficient foreground and background galaxies that the
projection corrections are well determined.  This redshift range
is also accessible to
other observational techniques which can be used to measure the mass profile,
in particular weak
gravitational lensing and X-ray observations,
although so far these are confined to the inner 1\hmpc\
of a cluster. These other observations will
provide constraints
on the relations between the virial temperatures
of the galaxy orbits, the X-ray gas, and the dark matter.
Moderate redshift clusters are also a superb match
to the observational capabilities of Canada France
Hawaii Telescope (CFHT), where a single
Multi-Object Spectrograph (MOS, Crampton \et 1993, LeF\`evre \et 1993,
Morbey, C. L. 1992)
field subtends $\simeq$2.5\hmpc\
and more than 100 galaxy spectra can be acquired in two hours at
the magnitude limit of our survey.

The volume density of high velocity dispersion clusters
is a sensitive indicator
of the amplitude of density fluctuations on cluster scales,
parameterized by $\sigma_8$, the fractional
mass variance in randomly placed spheres of 8\hmpc\ radius.
The bias parameter, $b=1/\sigma_8$, is essential for interpreting
the large scale flow velocities, and many other aspects of the
relations of luminous galaxies to the underlying dark mass fluctuations.
To overlap a measurement
of $\sigma_8$ with our mass profile study, we must select
a sample which is likely to have high velocity dispersions and
whose volume density can be well defined.
Accordingly, the clusters are selected to be an X-ray flux limited sample
with high luminosities.
The number density of clusters
is a strong function of the ``peak height'', which is directly
related to the cluster velocity dispersion, with a normalization
parameter which is
the amplitude of the density fluctuation spectrum on cluster
scales, $\sigma_8$. The
value of $\sigma_8$ is vigorously debated. It has been estimated
from optically selected cluster samples (Frenk \et 1990, White \et 1993)
and cluster X-ray luminosity and temperature functions
(\eg Evrard and Henry 1991, Kaiser 1991)
who argue that $\sigma_8\simeq 0.6$, but a somewhat different
theoretical calibration and allowance for redshift evolution in the sample
finds $\sigma_8$ in the 0.7 to 0.9 range (Bond and Myers 1992).
Our sample is effectively selected on the basis of the dark matter
potential well, and we only use the galaxies to gauge
the depth of the well. Our initial data allow us
to put an interesting new limit on $\sigma_8$.

This paper discusses the basic analysis which will be used
for cluster mass profiles and $\sigma_8$ measurements.
The reader is referred to Yee \et (1993) for a discussion
of the innovative observational techniques which ensure
statistically uniform selection of galaxies,
achieve a velocity accuracy of $\lta 150$ \kms as is
necessary for the study of cluster dynamics, and yields
10-20 $m_R\le 21.5$ redshifts per dark hour with the CFHT.

\goodbreak
\ls\ls\ni\ub{2. CLUSTER MASS PROFILES}

\ls
Luminosity segregation has long been recognized to be
a possibility within clusters, with the galaxies being more
concentrated to the center than the dark matter. There is
no evidence for or against such a segregation, given the relatively small
radial range of the data
(The and White 1986, Merritt 1987).
The discrepancy between cluster virial $\Omega$ values,
and the large scale streaming velocities $\Omega$ values,
which are measured
on scales only three times larger,
suggests that the excess mass
of $\Omega=1$ over $\Omega\simeq0.2$  may be located
on the outskirts of clusters (if it can undergo gravitational clustering).
The study of Coma by Kent and Gunn (1982)
had about 300 cluster velocities within
$3^\circ$ of the center from which they concluded that
a light-traces-mass model was entirely compatible with the
available data, but that a rising M/L could not be ruled out.
A test of whether M/L
rises with radius depends on getting many velocities
at as large an angular radius as possible, in which case
the projection of foreground and background objects into
the cluster's redshift space poses a severe problem.

The observational situation for Coma
is illustrated in Figure~1 using
ZCAT redshifts
(Huchra \et 1992)
for $m_B\le 16.5$ galaxies within
$15^\circ$ and 5000 \kms\ of the Coma cluster center.  This
dataset is a hodgepode, and {\it is not uniform across
the cluster} (van Haarlem \et 1993)
and will not give reliable masses. A more uniform sample
limited at $m_B=14.5$ does not have enough cluster galaxies
to be useful beyond about $2^\circ$.
The magnitude limit of 16.5 is
primarily chosen to give a sample size comparable to our survey.
The velocities have been
folded about the center of velocity and plotted against angular separation
from the center in order to average over the surrounding large
scale structure as much as possible. The average and RMS velocities
rise to nearly half of the 5000 \kms\ range,
indicating a severe field
contamination, even
at angles as small as $3^\circ$.

\setbox11=\vbox{
\ifapj\else \baselineskip 11pt \narrower \fi
\noindent
Figure 1:\hskip 5mm
The ZCAT distribution of velocities in the Coma field
($m_B\le 16.5$)
plotted against angular separation
folded about the central redshift, which helps to
smooth over surrounding large scale structure, and eliminates
linear selection gradients. The line gives the RMS
velocity in this background contaminated sample.
Note that the ZCAT sample is made from several heterogeneous
surveys of varying depth across the cluster, and
is far from the uniform survey which is essential
for mass estimates at large radii.
}
\ifapj \else
\midinsert
\epsfxsize 0.75\hsize
\epsfysize=4.0truein
\centerline{\epsffile{coma_vt.ps}}
\copy11 \endinsert \fi

\setbox12=\vbox{
\ifapj\else \baselineskip 11pt \narrower \fi
\noindent
Figure 2:\hskip 5mm
The background subtracted redshift map density distribution for the
Coma field. Contours of density are plotted at intervals
of 0.2 in log10, except for the bottom 5 contours which
are linear between 0.4 and $-$0.4. The bins are 500 \kms
by $1^\circ$. The solid line shows the estimated velocity
dispersion. Note the effective removal of
projected galaxies at large velocities.
Comparison to maps drawn from n-body data
indicates that this map is relatively free of problems to
a distance of $5^\circ$.
The errors in this map are dominated by the nonuniform sampling
of the cluster.
}
\ifapj \else \midinsert
\epsfysize=4.0truein
\epsfxsize 0.75\hsize
\centerline{\epsffile{coma_svt.ps}}
\copy12 \endinsert \fi

A simple approach to ``decontaminate'' the cluster redshift map
is to take advantage of the vast number of galaxies which
are in the surrounding field but which are
relatively unmoved by the cluster.
Under the assumption of uniform
sampling, the galaxy density at large radii can be
used to statistically estimate the contamination
through the cluster. For the illustrative
application to Coma the background
correction is based on the redshift map density
of galaxies at large angles,
$10^\circ$ to $15^\circ$.
For moderate redshift
clusters we will use the galaxies in front
and behind the cluster in redshift space. In this paper
the subtraction will
be done in bins of $[\Delta v, \theta]$, but the
same technique can be implemented within a maximum likelihood
model. The results of the field subtraction are shown in
Figure~2. Bins of 500 \kms and $1^\circ$ were used.
The velocity profile is not expected to be useful for
a serious mass estimate
because of non-uniform sampling of the sky amd
single clusters cannot give reliable mass profiles. Figure~2 does show
that the background subtraction technique qualitatively works
on real world data. The quantitative accuracy for
mass estimation is assessed with
simulation data.

The ability of the mass estimation method to discriminate between
low and high $\Omega$ universes
can be tested using results from n-body simulations. Two
simulations in 100\hmpc\ boxes are used,
one with $\Omega=0.2$, the other
with $\Omega=1$. In both cases the initial conditions
were $128^3$ particles perturbed with
a standard Cold Dark Matter (CDM) model evolved until $\sigma_8=1$ for
$\Omega=1$ and $\sigma_8=3$ (in this patch) for
$\Omega=0.2$. These parameters create a large cluster in each simulation
with a velocity dispersion of about 1000 \kms.
Samples of particles are drawn from the particles within
15\hmpc\ of the cluster center, and within 5000 \kms of
the line of sight velocity (including Hubble flow) of the
cluster center of velocity.
The $\Omega=0.2$ model is a light-traces-mass model, so
the sample of 2000 particles is drawn at random. To be realistic, the
tracers in the
$\Omega=1$ model must indicate a virial mass for the cluster
about 0.2 of the true virial mass.
Accordingly the sample
was created from a random sample of the
particles which were in $\rho\gta 1000\rho_0$ regions
at $z=3$.

\setbox13=\vbox{
\noindent
\ifapj\else \baselineskip 11pt \narrower \fi
\noindent
Figure 3:\hskip 5mm N-body tests of the contamination correction
and mass estimation for a light-traces-mass, $\Omega=0.2$ simulation
(left side plots) and a biased $\Omega=1$ simulation in which
the tracers indicate a cluster virial mass equivalent
to $\Omega_{VT}\simeq 0.2$
(right side plots). The plots are done for the
single largest cluster found in the
two simulations.
In the absence of cluster averaging there is no possibility
of distinguishing the difference $\Omega=0.2$, and $\Omega=1$ models,
because the velocity fields are dominated by local asymmetries.
More n-body simulations are planned (requiring about a cpu-year
of computing time on a modern workstation) to allow a representative
averaging over simulated clusters, but in the meantime the clusters
are averaged over the three independent projections (which does
not suppress local structure as well as 3 separate clusters would).
The simulation
data are scaled to the distance of Coma and similar numbers
of particles are used (about 400 in the cluster with a total of about 2000).
The top plots show corrected
density contours in the $[\Delta v,\theta]$ plane. The method
works well until the tracer density becomes comparable to the noise,
in which case characteristic ``islands'' and ``peninsulas'' appear.
In both cases the data are judged to be reliable to about $5^\circ$.
The bottom plots show the mass estimator. The dashed lines
indicate the enclosed light, which for the light-traces-mass
model on the left nicely traces the mass enclosed (shown by asterisks)
obtained by directly counting particles in spheres in the full
3D dataset.
}
\ifapj\else \midinsert
\epsfysize=6.0truein
\epsfxsize \hsize
\centerline{\epsffile{nbody.ps}}
\medskip
\copy13
\endinsert \fi

Figure 3 shows the corrected $[\Delta_v,\theta]$ maps
for the two simulations and the mass profiles,
real space measurements (asterisks) and estimated from
the decontaminated redshift map (triangles on the solid line).
The clusters are observed
from the three co-ordinate directions to help average
over the strong surrounding structure. The masses
are estimated using the projected mass estimator
(Bahcall and Tremaine 1981),
$$
M_P(<r_\theta) = {{S \sigma_1^2(r_\theta) r_\theta}\over{\pi G}},
\eqno{(1)}
$$
where $S$ is the orbit
shape coefficient which
ranges ranges from 16 to 32 for isotropic to completely radial
orbits.  The velocity ellipsoid shapes
are known from cosmological simulations
to be mildly radial, but as
a conservative choice we use $S=16$.
The values for $\sigma_1(\theta)$ are derived from the decontaminated
redshift map of Figure~2. To reduce graininess due to binning,
the RMS velocities
are accumulated for each bin, rather than simply being placed at the
center of the bin. The projected mass is normalized to
the virial mass, which is calculated
using all galaxies within 3000 \kms and $3^\circ$
of the center of the cluster,
$$
M_{VT} = {{3\pi\sigma_{1,RMS}^2}\over{2 G \langle{1/r_{\theta}}\rangle}}.
\eqno{(2)}
$$

The tests of the decontamination and mass estimation procedures using
simulation data shown in Figure~3 build confidence that
the method makes efficient use of the data to estimate cluster masses.
A dataset of 400 cluster velocities
and about 2000 total is sufficient to signal clearly the
presence or absence of extended dark halos,
out to a radius of about 6\hmpc.
It is also evident that the maps develop
``islands'' and ``peninsulas'' at large
$\Delta v$ and $r_p$, which are a clear indication
that the noise is overwhelming the signal. In the
$\Omega=0.2$ model of Figure~3 the cleaned integrated light distribution
does a remarkable job of following the integrated mass
distribution, indicating that the decontamination
works extremely well on integral quantities. Most importantly,
for both the $\Omega=1$ and $\Omega=0.2$ datasets
the estimated mass profile is close to the true mass
distribution. The radial bins are $1^\circ$ here, although averaging
over $2^\circ$ gives a somewhat improved $M(<r)$ estimate, particular for
the $\Omega=0.2$ case.
Because the three projections are not independent,
the effective amount of data will be somewhat smaller than
the $3\times 400$ cluster particles selected in each case.
It appears safe to conclude that with a sample of about 1000 cluster
galaxies, spread over a number of clusters, it will be possible
to measure the mass profile of the cluster, with a
statistical accuracy nearly at the
$\sqrt{N}$ level of the amount of data at large radii.

Simulating only two clusters serves to
illustrate the effectiveness of the method, but it does not give an
indication of the errors.
Ultimately a sample of 100 or so
realizations of these two models will be studied to
better determine the errors of mass determination, estimate
the effect of substructure in velocity distributions
on velocity dispersion estimates, and to calibrate the
relation between
observed velocity dispersion and the volume density of
clusters, as discussed below using an analytic theory.

\ls\ls\ifapj \else\par\penalty-10000\fi
\ni\ub{3. THE DENSITY FLUCTUATION SPECTRUM AMPLITUDE}

\ls
The amplitude of the density fluctuation spectrum,
$P(k)$, is conventionally given by $\sigma_8$, defined as the linear
extrapolation of the mass variance in a sphere of 8\hmpc\
radius measured at the current epoch.
In the galaxy
population the variance of number density is equal to 1,
within the errors; hence the bias parameter is
$b=1/\sigma_8$. For the  ``classical'' CDM spectrum,
$\sigma_8\simeq 0.4$ (Davis \et 1985) if $\Omega=1$
and galaxies accurately represent the pairwise velocities of
the dark matter population. The large scale streaming
velocities are so much higher than those which
a $\sigma_8=0.4$ model predicts,
that such a low $\sigma_8$ has long been ruled out. Furthermore,
the large scale streaming velocities
indicate $\Omega^{0.6}/b\simeq 1$; so that a knowledge of
$\sigma_8$ can be combined into an estimate of $\Omega$.
The COBE measurement of temperature fluctuations
in the cosmic background radiation  (Smoot \et 1992),
when extrapolated with the $n=1$ CDM
spectrum, predicts that $\sigma_8\simeq 1.2$ (Wright \et 1992,
for a quadrupole
amplitude of 15$\mu$K); therefore measurement of
$\sigma_8$ sets a strong constraint on the shape of the spectrum.

Clusters contain a small fraction of the mass of the
universe, meaning that they form from relatively rare perturbations
in the extreme tail of the probability distribution.
Therefore
the number of clusters per unit volume is a
sensitive indicator of the amplitude of the perturbation
spectrum. As shown below, a factor of two variation
in $\sigma_8$ changes the number density of clusters by two
orders of magnitude. Clusters offer the considerable benefit
that they can be viewed as a direct sample of the dark matter,
with galaxies simply being a thermometer of the cluster velocity
dispersion. Much of the argument over clusters and the density
fluctuation spectrum comes from optically selected clusters,
which are subject to projection effects which artificially
increase the apparent galaxy richness of clusters.

Clusters can
be selected from X-ray surveys, which are nearly
ideal for our purposes, in that objects are found on the basis
of the depth of the potential well
and projection overlap of clusters in X-rays is small.
The EMSS serendipitous X-ray survey
of Gioia \et (1990, Henry \et 1992)
has a low enough flux limit to find many luminous clusters
at moderate redshifts.
Of particular interest are the most luminous clusters, those
with $L_x>4\times10^{44}$ erg s$^{-1}$, for which several
X-ray luminosity function studies have claimed a strong
evolution in numbers between
$z=0$ and $z\simeq 0.3$ (Edge \et 1990,
Henry \et 1992). Moreover such
clusters are are most likely to have a high velocity dispersion.
Selecting clusters with $z>0.18$, $L_x>4\times 10^{44}$ erg~s$^{-1}$,
$f_x>5\times10{-13}$ erg~s$^{-1}$~cm$^{-2}$ (to ensure inclusion
in the ROSAT survey), and in the
declination range of $-$5 to 40$^\circ$ (convenient for
long CFHT exposures)
gives the sample in Table~1.

\ifapj\baselineskip 15pt\fi
\setbox21= \vbox{\parindent 0pt
{Table 1: The Cluster Sample}
\bigskip
\offinterlineskip
{\halign{
\vrule \vrule height 8.5pt depth3.5pt width0pt #& \
\quad #\quad\hfil &\vrule#&
\quad\hfil #\quad &\vrule#&
\quad\hfil #\quad &\vrule#&
\quad\hfil #\quad &\vrule# \cr
\noalign{\hrule}
& Name && redshift && $L_x/10^{44}$ && $f_x/10^{-13}$ &\cr
\noalign{\hrule}
&MS0015.9 +1609 && 0.540 &&  14.31 &&  11.60  &\cr
&MS0302.7 +1658 && 0.424 &&  4.99 &&  6.54  &\cr
&MS0440.5 +0204 && 0.190 &&  4.00 &&  25.92  &\cr
&MS0451.5 +0250 && 0.202 &&  6.96 &&  39.92  &\cr
&MS0451.6 $-$0305 && 0.547 && 25.10 &&  10.75  &\cr
&MS0839.8 +2938 && 0.194 &&  5.33 &&  33.17  &\cr
&MS0906.5 +1110 && 0.180 &&  5.75 &&  41.56  &\cr
&MS1006.0 +1202 && 0.221 &&  4.80 &&  23.04  &\cr
&MS1224.7 +2007 && 0.327 &&  4.59 &&  10.08  &\cr
&MS1333.3 +1725 && 0.460 &&  5.39 &&  6.00  &\cr
&MS1455.0 +2232 && 0.259 &&  15.98 &&  55.85  &\cr
&MS1512.4 +3647 && 0.372 &&  4.80 &&  8.15  &\cr
&MS1621.5 +2640 && 0.426 &&  4.52 &&  5.87  &\cr
\noalign{\hrule}
}
}}
\ifapj\baselineskip 24pt\fi
\ifapj\else \midinsert \copy21 \endinsert \fi

The expected volume density of clusters as a function of $\sigma_8$ can
be calculated using the Press-Schechter (1974) theory, which has
been shown to be an excellent predictor of the numbers of
halos found in n-body simulations (Efstathiou \et 1988,
Carlberg and Couchman 1989). The number of
halos in the range of velocity dispersion
$\sigma_v$ to $\sigma_v+ d\sigma_v$ is,
$$
n(\sigma_v) d\sigma_v = {{9c_v^3 H_0^3}\over{4\pi\sqrt{2\pi}}}
	{(1+z)^{3/2}\over{\sigma_v^4}}
	{{d \ln{\sigma(M)}\over{d\ln{\sigma_v}}}}
	\nu e^{-\nu^2/2} d{\sigma_v},
\eqno{(3)}
$$
where $\nu=1.68(1+z)/(\sigma(M)\sigma_8)$ and we have
assumed that $\Omega=1$. The
mass variances, $\sigma(M)$, are calculated from
the CDM spectrum. Masses are related
to line of sight velocity dispersions as $\sigma_v=c_v H_0 R(1+z)^{1/2}$,
where $M=4\pi/3\rho_0R^3$.
The constant $c_v$ is not well defined, and will be the
subject of a large Monte Carlo n-body study. The spherical collapse
of a tophat sphere gives $c_v=1.18$ (White and Frenk 1991).
However, examination of the largest collapsed peaks in the n-body simulations
gives $c_v=1.10$, which we will adopt (see also Bond and Myers 1993).
The tracer population in the cluster of
the $\Omega=1$ model has a considerably lower velocity
dispersion, corresponding to $c_v=0.85$, but this
single particle velocity bias is best handled in
conjunction with an indication of the mass profile
at large radii. A conservative value for the likely
velocity bias in clusters is adopted, $b_v=0.9$, that
is, a 10\% reduction of the galaxy RMS velocity from
the dark matter values (Carlberg and Dubinski 1991).

The solid angle covered by the EMSS sample above our flux limit
and in our range of declination is estimated to
be 223 square degrees (Henry \et 1992), and the
total co-moving volume between $z=0.18$ and $z=0.54$ is
$3.3\times10^7$ h$^3$~Mpc$^3$, for $q_0=0.5$. Lower $q_0$ give
larger volumes.
The range of densities
is estimated on the basis that there is at least one
cluster in this volume, and no more than 12.
The predicted volume densities for all clusters
with dispersions greater than a given value is the
integral of Equation (3) from $\sigma_v$ to infinity.

\ifapj\baselineskip 15pt\fi
\setbox22= \vbox{\parindent 0pt
{Table 2: Cluster Velocity Dispersions}
\bigskip
\offinterlineskip
{\halign{
\vrule \vrule height 8.5pt depth3.5pt width0pt #&\quad #\quad\hfil&\vrule#&
\quad\hfil #\quad &\vrule#&
\quad\hfil #\quad &\vrule#&
\quad\hfil #\quad &\vrule#&
\quad\hfil #\quad &\vrule# \cr
\noalign{\hrule}
& Name && redshift && $N_c$ && $\sigma_v$ && 68\% confidence &\cr
\noalign{\hrule}
& MS0451.5 +0250 && 0.2015 &&  24 &&  1097 && 984-1252  &\cr
& MS0839.8 +2938 && 0.1924 &&  32 &&  939  && 796-1071 &\cr
& MS1224.7 +2007 && 0.3250 &&  29 &&  837  && 756-931  &\cr
\noalign{\hrule}
}
}}
\ifapj\baselineskip 24pt\fi
\ifapj\else \midinsert \medskip \copy22 \medskip \endinsert \fi

\setbox14=\vbox{
\ifapj\else \baselineskip 11pt \narrower \fi
\noindent
Figure 4:\hskip 5mm
The volume density of clusters greater than a specified velocity
dispersion at $z=0.3$. The theoretical curves are labeled with
their normalizing $\sigma_8$ values. Note that a factor of
two change in $\sigma_8$ gives two decades of change in the
volume density. The data constraints shown are preliminary
results based on three clusters with 30 redshifts each.
The lower limit is for the volume enclosed by the highest
redshift cluster, MS1224+20, plotted with $b_v=1$.
The error box
lower limit assumes one cluster with a
dispersion of 950 \kms in
the redshift range 0.18 to 0.54; the upper limit assumes all
12 are in this range. The lower limit
uses $b_v=1$, and the combined data use $b_v=0.9$.
}
\ifapj\else \midinsert
\epsfysize=4.0truein
\epsfxsize 0.75\hsize
\centerline{\epsffile{nofv.ps}}
\medskip
\copy14
\endinsert \fi

Preliminary velocity dispersions of three of the clusters are reported
in Table~2, and have been
subjected to a variety of statistical tests
(Bird and Beers 1993) for credibility.
Eventually we will obtain about 150 velocities per
cluster, thereby greatly improving the statistical accuracy of the
dispersions, and allowing clear separation of
neighbouring groups.
The cluster
MS0839+29 has a gap of nearly 600 \kms in its velocity
histogram, which is statistically significant in comparison
to a Gaussian (which cluster velocity distributions are
not expected to be).
Eliminating
this gap, and a second smaller one which becomes significant
after the first velocity substructure is dropped,
reduces the velocity dispersion
to about 400 \kms for this cluster, which would be unusually
small, given its X-ray luminosity (Edge and Stewart 1992).
Dropping this cluster from the sample would make little difference
to either the volume or the average velocity dispersion displayed
in Figure~5. The highest redshift cluster can be used to
put a lower limit on the Figure~5. The cluster MS1224+20
has a velocity dispersion of 840 \kms at $z=0.325$. The co-moving volume
from $z=0.18$ is $1.3\times10^8 h^{-3}$ Mpc$^3$.
The lower limit
point is plotted with no allowance for velocity bias, which is
inconsistent with clusters having extended dark matter halos.

The implications of this small sample for $\sigma_8$ are
shown in Figure~5. The three clusters have been averaged
and
$b_v=0.9$ is adopted. The lower limit is for the single highest redshift
cluster, MS1224+20, which has a velocity dispersion of 840 \kms
with the volume is that between our lowest allowed
redshift $z=0.18$ and the cluster redshift, $0.327$.
We tentatively find
that the value of $\sigma_8$ is
constrained to be in the range of 0.6 to 0.9, which is somewhat
higher than found by the studies of the evolution
of the X-ray luminosity function
function (Evrard and Henry 1991, Kaiser 1991, Henry \et
1992) but similar to Bond and Myers (1992).
Our $\sigma_8$ value combined with the COBE result (Wright \et 1992)
is sufficient to argue that either
$P(k)$
must have relatively more long wave power than CDM, as suggested
by IRAS galaxy surveys (Saunders \et 1991),
or that there are extra sources of variance in the
cosmic background radiation (such
as primordial gravitational waves).
The most significant
aspect of this result is a relatively convincing demonstration,
based entirely on X-ray selection and velocity dispersion
measurements, that
the value of $\sigma_8$ is in the range of 0.6 to 0.9, and therefore
the bias factor, $b$, is 1.6 to 1.1. When our result is
combined with the streaming velocity result, $\Omega^{0.6}/b\simeq 0.7$
(Kaiser \et 1991, Strauss \et 1992), it implies
that the large scale value
of $\Omega$ is most likely about 0.5(!). We emphasize that this
is a preliminary analysis based on a small dataset over a limited
redshift range which will be substantially strengthened when our
entire dataset is in hand.

\ls\ls\goodbreak\ni\ub{4. CONCLUSIONS}

\ls
The CNOC cluster mapping project is designed to provide
extensive dynamical
data for a small sample of high X-ray luminosity clusters around $z\simeq0.3$.
The galaxies are observed with a technique of high velocity accuracy
for the redshift range. The total random
velocity errors are $\lta$150 \kms of which only 100 \kms is due to
velocity estimation, the rest being due to slit position and
wavelength calibration uncertainties which potentially
can be greatly reduced. The primary use of
the data are to search for extended dark matter halos around
clusters, with a secondary application being to measure the value
of the $\sigma_8$ normalization parameter for the amplitude
of the density perturbation spectrum on cluster scales.

The data in hand, from a single telescope run of three clear nights
in January 1993, are sufficient to put a new, fairly tight, limit
on $\sigma_8$, confining it to the range of 0.6 to 0.9, using
the volume density of high velocity dispersion clusters.
The result is based on a cluster sample selected from an X-ray
survey, which is insensitive to projection effects in sample
construction and appears to be successful for finding
rich, high velocity dispersion clusters.

A simple
background decontamination technique has been developed which
allows us
to measure the extended mass profiles of clusters.
The method
has been  applied to small samples of simulated data
drawn from n-body simulations. For about a 1000 cluster galaxies
spread over half a dozen clusters, the method is sufficient
to distinguish been the cluster M/L
profiles characteristic of  $\Omega=0.2$ and
$\Omega=1$ universes,
at a confidence level of 2-3 standard
deviations.

At the present time we have about 230 redshifts, about 10\% of
the number that we need to complete this project to the accuracy
necessary to get a statistically reliable result. The intent
of this paper is to give a preliminary discussion of the overall
goals of the project, the basic analysis procedures, and some
tentative results.

\bigskip
\ni
{\bf Acknowledgements:} We thank the CFHT
Canadian Time Allocation Committee
for a grant of
observing time. CFHT staff, in particular TO's  Ken Barton,
John Hamilton, and Norm Purvis,
provided invaluable assistance
at the telescope. The MOS team is to be congratulated
for creating a superb instrument.
Financial support for this project comes
from the Natural Sciences and Engineering Research Council of
Canada, the National Research Council, the Canadian Institute
for Advanced Research, the Science and Engineering
Research Council of the United Kingdom,
the National Science Foundation and the National Aeronautics and
Space Administration of the
United States of America, and the Counsel Nationale de R\'echerche
Scientifique of France.

\ifapj \vfill\eject \fi
\goodbreak
\ls\ls\ls
\ni\ub{REFERENCES}
\parskip=0pt

\ls
\hi{Bahcall, J. \& Tremaine, S. D. 1981, \apj{244}, 805}
\hi{Bardeen, J. M., Steinhardt, P. J. \& Turner, M. S. 1983, Phys. Rev. D., 28,
679}
\hi{Bean, A. J., Efstathiou, G., Ellis, R. S., Peterson, B. A. \& Shanks, T.
1983, \mn 205, 605}
\hi{Bertschinger, E., Dekel, A., Faber, S. M., Dressler, A. \& Burstein, D.
1990, \apj  364, 370}
\hi{Bird, C. M. \& Beers, T. C. 1993, \aj 105, 1596} 
\hi{Bond, J. R. \& Myers, S. T. 1992, in {\it Trends in Astroparticle
	Physics} (World Scientific: Singapore), p. 262}
\hi{Bond, J. R. \& Myers, S. T. 1993, \apj submitted}
\hi{Carlberg, R. G. \& Dubinski, J. 1991, \apj 369, 13}
\hi{Carlberg, R. G. \& Couchman, H. M. P. 1989, \apj 340, 47}
\hi{Carlberg, R. G., Couchman, H. M. P. \& Thomas, P. A. 1990, \apjl {\bf 352},
L29}
\hi{Crampton, D. \et 1993, in {\it Proc. ESO Conference on Progress
	in Telescope and Instrumentation Technologies}, ed. M.-H. Ulrich
	(ESO: Garching) in press}
\hi{Davis, M. \& Peebles, P. J. E. 1983, \apj {267}, 465}
\hi{Davis, M., Efstathiou, G., Frenk, C. S. \& White, S. D. M. 1985, \apj 292,
371}
\hi{Edge, A. C., Stewart, G. C., Fabian, A. C., Arnaud, K. A. 1990,
	\mn 245, 559} 
\hi{Edge, A. C. \& Stewart 1992, \mn 252, 428} 
\hi{Efron, B. \& Tishirani, R. 1986, {\it Statistical Science}, 1, 54}
\hi{Efstathiou, G., Frenk, C. S., White, S. D. M., \& Davis, M.
        1988, \mn  235, 715}
\hi{Evrard, A. E. \& Henry, J. P. 1991, \apj{383}, 95}
\hi{Frenk, C. S., White, S. D. M., Efstathiou, G. E., \& Davis, M 1990,
	\apj 351 10}
\hi{Gioia, I. M., Maccacaro, T., Schild, R. E., Wolter, A., Stocke, J. T.,
	Morris, S. L., \& Henry, J. P. 1990, \apjs 72, 567}
\hi{Guth, A. 1981, Phys. Rev. D., 23, 347}
\hi{Henry, J. P., Gioia, I. M., Maccacaro, T., Morris, S. L.,
	Stocke, J. T., \& Wolter, A. 1992, \apj 386, 408}
\hi{Huchra, J. P., Geller, M. J., Clemens, C. M., Tokarz, S. P.,
	Michel, A. 1992,
	{\it The CfA Redshift Catalogue, ``ZCAT''},
	Bull. Inf. CDS 41, NASA NSSDC}
\hi{Kaiser, N. 1991, \apj 383, 104}
\hi{Kaiser, N., Efstathiou, G., Ellis, R., Frenk, C., Lawrence, A.,
Rowan-Robinson, M., \& Saunders, W. 1991, \mn 252, 1}
\hi{Kent, S. \& Gunn J. E. 1982, \aj 87, 945}
\hi{LeF\`evre, O., Crampton, D., Felenbok, P., \& Monnet, G. 1993, preprint}
\hi{Merritt, D. 1987, \apj 313,  121}
\hi{Morbey, C. L. 1992, {\it Applied Optics}, 31, 2291}
\hi{Peebles, P. J. E. 1971, {\it Physical Cosmology}, Princeton University
	Press}
\hi{Peebles, P. J. E. 1980, {\it The Large-Scale Structure of the
	Universe}, Princeton University Press}
\hi{Peebles, P. J. E. 1993, {\it Principles of Physical Cosmology},
	Princeton University Press}
\hi{Press, W. H. \& Schechter, P. 1974, \apj 187, 425}
\hi{Saunders, W., Frenk, C. S., Rowan-Robinson, M., Efstathiou, G., Lawrence,
A., Kaiser, N., Ellis, R., Crawford, J., Xia, X.-Y. \& Parry, I. 1991, Nature,
{\bf 349}, 32} 
\hi{Smoot, G. F., Bennett, C. L., Kogut, A., Wright, E. L., Aymon, J., Boggess,
N. W., Cheng, E. S., De Amici, G., Gulkis, S., Hauser, M. G., Hinshaw, G.,
Jackson, P. D., Jannsen, M., Kaita, E., Kelsall, T., Keegstra, P., Lineweaver,
C., Lowenstein, K., Lubin, P., Mather, J., Meyer, S. S., Moseley, S. H.,
Murdock, T., Rokke, L., Silverberg, R. F., Tenorio, L., Weiss, R., \&
Wilkinson, D. T.  1992, \apjl 326, L1}
\hi{Strauss, M. A., Yahil, A., Davis, M., Huchra, J. P., \& Fisher, K. 1992,
\apj 397, 395}
\hi{The, L. S. \& White, S. D. M. 1986, \aj 92, 1248}
\hi{van Haarlem, M., Coyon, L., Guitierrez de la Cruz, C., Martinez-Gonzalez,
	E., \& Rebolo, R. 1993, \mn 264, 71}
\hi{West, M. J., \& Richstone, D. O. 1988, \apj 335, 532}
\hi{White, S. D. M. \& Frenk, C. S. 1991, \apj{379}, 52}
\hi{White, S. D. M., Efstathiou, G., \& Frenk, C. S. 1993, \mn 262, 1023}
\hi{Wright, E. L., Meyer, S. S., Bennett, C. L., Boggess, N. W., Cheng, E. S.,
Hauser, M. G., Kogut, A., Lineweaver, C., Mather, J. C., Smoot, G. F., Weiss,
R., Gulkis, S., Hinshaw, G., Jannssen, M., Kelsall, T., Lubin, P. M., Moseley,
S. H. Jr., Murdock, T. L., Shafer, R. A., Silverberg, R. F., \& Wilkinson, D.
T. 1992, \apjl 396, L13.}
\hi{Yee, H. K. C., \et 1993, in preparation}

\ifapj
\vfill \eject
\medskip
\copy21
\bigskip
\copy22
\bigskip \bigskip \bigskip
\centerline{FIGURE CAPTIONS}
\medskip
\copy11
\bigskip
\copy12
\bigskip
\copy13
\bigskip
\copy14
\bigskip
\copy15
\vfill \eject
\centerline{\caps Fig. 1}
\bigskip
\epsfysize=7.5truein
\centerline{\epsffile{coma_vt.ps}}
\vfill\eject
\centerline{\caps Fig. 2}
\epsfysize=7.5truein
\centerline{\epsffile{coma_svt.ps}}
\vfill\eject
\centerline{\caps Fig. 3}
\epsfysize=7.5truein
\centerline{\epsffile{nbody.PS}}
\vfill \eject
\centerline{\caps Fig. 4}
\epsfysize=7.5truein
\centerline{\epsffile{nofv.PS}}
\vfill \eject
\fi

\bye